\documentclass[a4paper,11pt]{article}

\usepackage{amssymb,amsmath,amsthm,fullpage}
\usepackage{tikz}
\usepackage[pdftex]{hyperref}
\hypersetup{plainpages=True, pdfstartview=FitH, bookmarksopen=true, colorlinks=true,linkcolor=blue,citecolor=blue}

\def\pf{\noindent \emph{Proof.}\ }
\def\qed{{\quad\rule{1mm}{3mm}\,}}

\newtheorem{theorem}{Theorem}

\newtheorem{corollary}{Corollary}
\newtheorem{lemma}{Lemma}
\theoremstyle{remark}\newtheorem{remark}{Remark}
\newtheorem{example}{Example}
\newtheorem{definition}{Definition}

\begin{document}
\title{A Short Note on the Exact Counting of Tree-Child Networks}
\author{Michael Fuchs\\
Department of Mathematical Sciences\\
National Chengchi University\\
Taipei 116\\
Taiwan
\and Hexuan Liu\\
Department of Mathematical Sciences\\
National Chengchi University\\
Taipei 116\\
Taiwan
\and Guan-Ru Yu\\Department of Mathematics\\
National Kaohsiung Normal University\\
Kaohsiung 824\\
Taiwan}
\date{\today}
\maketitle

\begin{abstract}
Tree-child networks are an important network class which are used in  phylogenetics to model reticulate evolution. In a recent paper, Pons and Batle (2021) conjectured a relation between tree-child networks and certain words. In this short note, we prove their conjecture for the (important) class of one-component tree-child networks.
\end{abstract}

\section{Introduction}

Due to mechanisms such as hybridization, lateral gene transfer or recombination, many models in evolutionary biology are nowadays based on phylogenetic networks rather than phylogenetic trees. Moreover, in order to make phylogenetic networks amenable to methods from bioinformatics,  topological constraints have been considered for these networks which lead to the introduction of many different classes of phylogenetic networks; see \cite{HuRuSc}. One of the most popular of these classes is the class of tree-child networks. Consequently, several recent studies have investigated the properties of this class; see \cite{CaZh, FuYuZh, PoBa} and references therein.

We first recall the definition of tree-child networks.

\begin{definition}
A {\it tree-child network} with $n$ leaves is a rooted DAG without double edges such that all nodes belong to one of the following categories:
\begin{itemize}
\item[(i)] A unique {\it root} which has indegree $0$ and outdegree $1$.
\item[(ii)] {\it Leaves} which have indegree $1$ and outdegree $0$ and are labeled bijectively with labels from the set $\{1,\ldots,n\}$.
\item[(iii)] {\it Tree nodes} which have indegree $1$ and outdegree $2$.
\item[(iv)] {\it Reticulation nodes} which have indegree $2$ and outdegree $1$.
\end{itemize}
In addition, each internal node satisfies the {\it tree-child property}: it has at least one child which is not a reticulation node.

We denote by $\mathcal{TC}_{n,k}$ the set of tree-child networks with $n$ leaves and $k$ reticulation nodes.
\end{definition}

\begin{remark}
Recall that for the the number $k$ of reticulation nodes, we have that $0\leq k\leq n-1$, where for $k=0$, the set $\mathcal{TC}_{n,0}$ is the set of {\it phylogenetic trees}; see, e.g., \cite{SeSt}.
\end{remark}

Several recent papers have investigated counting questions for tree-child networks; see \cite{CaZh,FuYuZh,PoBa}. One fruitful observation was that tree-child networks can be encoded by words.

To give details, we consider the following set from \cite{PoBa}.

\begin{definition}
Let $\mathcal{C}_{n,k}$ denote the class of words on $n$ letters  $\{\omega_1,\ldots,\omega_{n}\}$ such that $k$ letters occur $3$ times and $n-k$ letters occur $2$ times. Moreover, these words satisfy that in every prefix, the letter $\omega_i$ either has not yet occurred or if it has, then the number of occurrences is not smaller than the number of occurrences of $\omega_j$ for all $j>i$.
\end{definition}

Using this set, in \cite{FuYuZh}, the authors proved that for tree-child networks with a maximal number of reticulation nodes:
\[
\vert\mathcal{TC}_{n,n-1}\vert=n!\times\vert\mathcal{C}_{n-1,n-1}\vert.
\]
Moreover, in \cite{PoBa}, the authors obtained a similar relation as above for phylogenetic trees:
\begin{equation}\label{result-bp}
\vert\mathcal{TC}_{n,0}\vert=\vert\mathcal{C}_{n-1,0}\vert.
\end{equation}

These two extreme cases inspired the authors of \cite{PoBa} to pose the following conjecture:
\[
\vert\mathcal{TC}_{n,k}\vert=\frac{n!}{(n-k)!}\times\vert\mathcal C_{n-1,k}\vert,\qquad (0\leq k\leq n-1).
\]
This conjecture is consistent with previous results about the counting of tree-child networks and has several interesting consequences; see \cite{PoBa}.

The purpose of this short note is to verify the above conjecture for one-component tree-child networks which are an important building block of (general) tree-child networks; see \cite{CaZh}.

\begin{definition}
A tree-child network is called {\it one-component tree-child network} (OTC) if every reticulation node is directly followed by a leaf. We denote by $\mathcal{OTC}_{n,k}$ the class of OTCs with $n$ leaves and $k$ reticulation nodes.
\end{definition}

For this class, we will consider the following subset of $\mathcal{C}_{n,k}$

\begin{definition}
Let  $\mathcal{H}_{n,k}$ denote the subset of words from $\mathcal{C}_{n,k}$ which satisfy that the letters before the third occurrence of a letter are the second occurrence of these letters.
\end{definition}

\begin{example}
An example for a word in $\mathcal{H}_{5,2}$ with the letters in the order $a,b,c,d,e$ is $ebadacbacdeb$.
\end{example}

We can now formulate our main result, where we use the notation $A^{\underline{k}}$ for the set consisting of $k$-tuples without repetition of elements from $A$.

\begin{theorem}\label{main-result}
There is a bijection between $\mathcal{OTC}_{n,k}$ and $\mathcal{H}_{n-1,k}\times\{1,\ldots,n\}^{\underline{k}}$. Thus,
\[
\vert\mathcal{OTC}_{n,k}\vert=\frac{n!}{(n-k)!}\times\vert\mathcal{H}_{n-1,k}\vert.
\]
\end{theorem}

The proof of this result is contained in the next section; the last section contains concluding remarks.

\section{Proof of the Main Result}

We start with some (simple) facts.

First, the letters which occur three times in a word from $\mathcal{C}_{n,k}$ must be the letters $\omega_1,\ldots,\omega_k$ and the third occurrence of these words must happen in this order.

Next, we state a simple lemma.

\begin{lemma}\label{hnk}
There is a bijection between $\mathcal{H}_{n,k}$ and $\mathcal{C}_{n,0}\times\binom{\{1,\ldots,n\}}{k}$, where $\binom{A}{k}$ denotes the set consisting of subsets of $A$ of cardinality $k$.
\end{lemma}

\pf Given a word from $\mathcal{H}_{n,k}$ first remove the third occurrence of the letters of $\omega_1,\ldots,\omega_k$. Note that the resulting word belongs to $\mathcal{C}_{n,0}$. Moreover, this word and the set of letters preceding $\omega_1,\ldots,\omega_k$ allow us to reconstruct the original word.\qed

We will in the sequel identify words from $\mathcal{H}_{n,k}$ with the image under the above map, where we use the notation from the next example.

\begin{example}\label{not-words}
For the example from the previous section, we have that the image under the above map consists of the word $ebadacbcde$ and the set $\{b,e\}$. We will write this in the sequel as $ebadacbce+\{b,e\}$.
\end{example}

We can now prove Theorem~\ref{main-result}. We will show how to construct the map and its inverse between the two sets in the next two paragraphs.

\vspace*{0.35cm}\noindent
{\bf From  $\mathcal{OTC}_{n,k}$ to  $\mathcal{H}_{n-1,k}\times\{1,\ldots,n\}^{\underline{k}}$.}

\begin{figure}[t!]
\begin{center}
\includegraphics[scale=0.68]{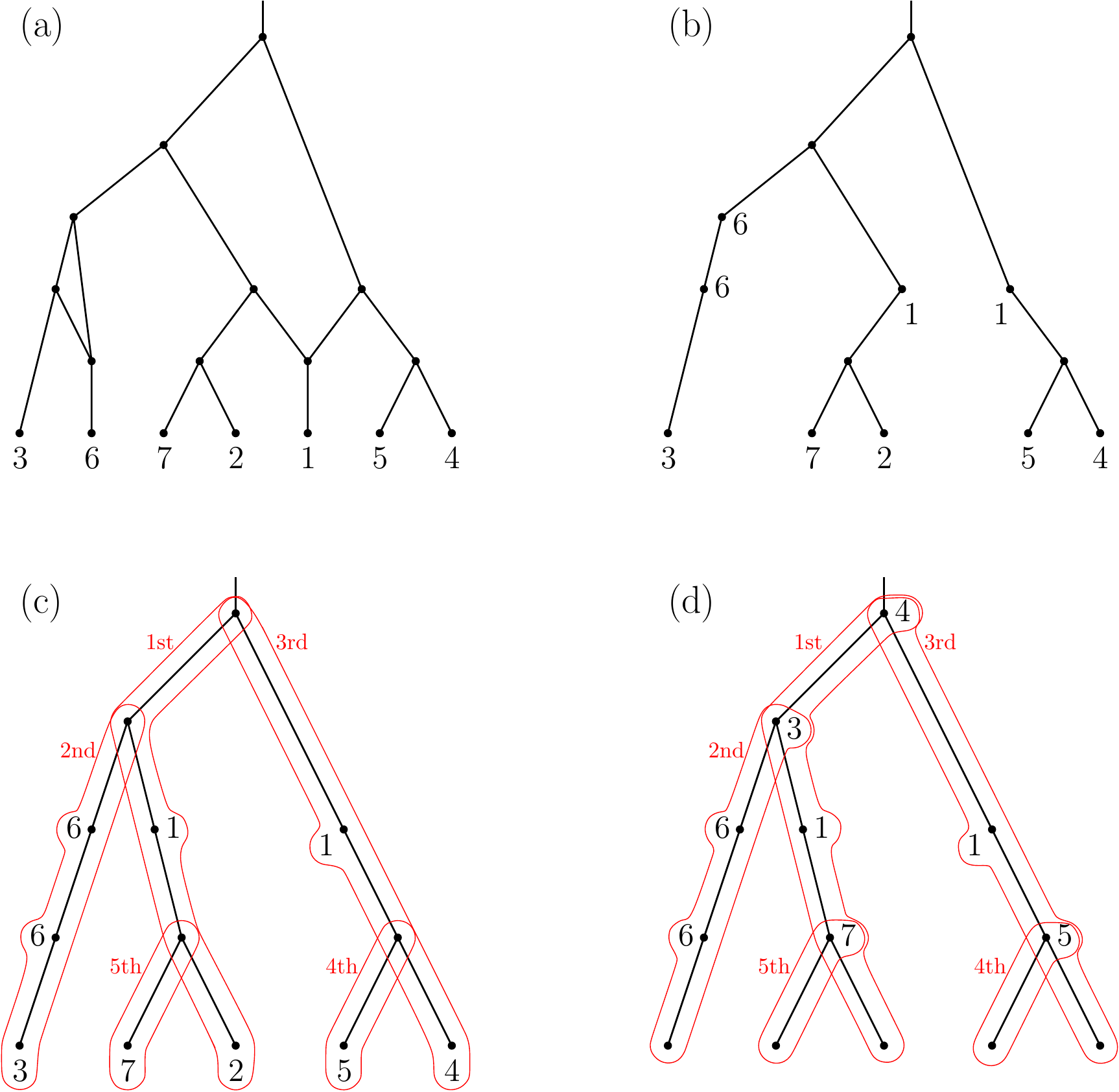}
\end{center}
\caption{(a) An example of an OTC $N$; (b) $N$ with the reticulation nodes removed which yields a tree $T$; (c) the path decomposition of $T$; (d) $T$ with internal nodes labeled and labels of leaves removed.}\label{path-dec}
\end{figure}

\vspace*{0.35cm} Assume that a OTC $N$ from $\mathcal{OTC}_{n,k}$ is given; see Figure~\ref{path-dec}-(a).

We first remove the reticulation nodes, the edges and leaves below them and their two incoming edges. Moreover, we label their parents (which are now unary nodes, i.e., nodes of indegree $1$ and outdegree $1$) with the labels of the (removed) leaves below the reticulation nodes; see Figure~\ref{path-dec}-(b).

Next, we decompose the resulting tree into paths where each path has an index: the first one is the (unique) path from the leaf with the smallest label to the root; next, find the leaf with the second smallest label; the second path is from that leave to the node which lies already on an indexed path; etc; see Figure-\ref{path-dec}-(c).

Using the above path decomposition, we label the (remaining) internal nodes as follows: the first node on each path is labeled by the label of its leaf. Moreover, we remove the labels of the leaves; see Figure-\ref{path-dec}-(d).

Next, we read the labels on the indexed paths starting with the first one, then the second one, etc. This gives a word $\tilde{\omega}$. Also put the labels of the unary nodes into a set $\tilde{A}$. For the example from Figure~\ref{path-dec}, this gives (with the notation from Example~\ref{not-words}): $431736641557+\{1,6\}$.

From $\tilde{\omega}$ and $\tilde{A}$, the image of $N$ under our bijection is now constructed as follows.

First, construct a $k$-tuple of integers by recording the second occurrences of the letters of $\tilde{A}$ in $\tilde{\omega}$ in the order in which they occur; for the example from Figure~\ref{path-dec}, we have $(6,1)$.

Second, we rename the numbers in $\tilde{\omega}$ and $\tilde{A}$ as follows: if a number in $\tilde{\omega}$ appears the second time and there are already $j-1$ numbers appearing twice before it, then rename the number by $\omega_{j}$. This converts  $\tilde{\omega}$ and $\tilde{A}$ into $\omega$ and $A$. Moreover, via the map from Lemma~\ref{hnk}, this gives a word from $\mathcal{H}_{n-1,k}$. For the example from Figure~\ref{path-dec}, the mapping is:
\[
3\rightarrow a,\quad 6\rightarrow b,\quad 4\rightarrow c,\quad 1\rightarrow d,\quad
5\rightarrow e,\quad 7\rightarrow f.
\]
Thus, $\omega=cadfabbcdeef$ and $A=\{b,d\}$. Finally, this gives the word: $cadfabbacdbeef$.

The above word  and the $k$-tuple are the image of $N$ under our bijection. For the example from Figure~\ref{path-dec}, the image is: $cadfabbacdbeef$ and $(6,1)$.

\vspace*{0.35cm}\noindent\textbf{From $\mathcal{H}_{n-1,k}\times\{1,\ldots,n\}^{\underline{k}}$ to $\mathcal{OTC}_{n,k}$.}

\begin{figure}[t!]
\begin{center}
\includegraphics[scale=0.75]{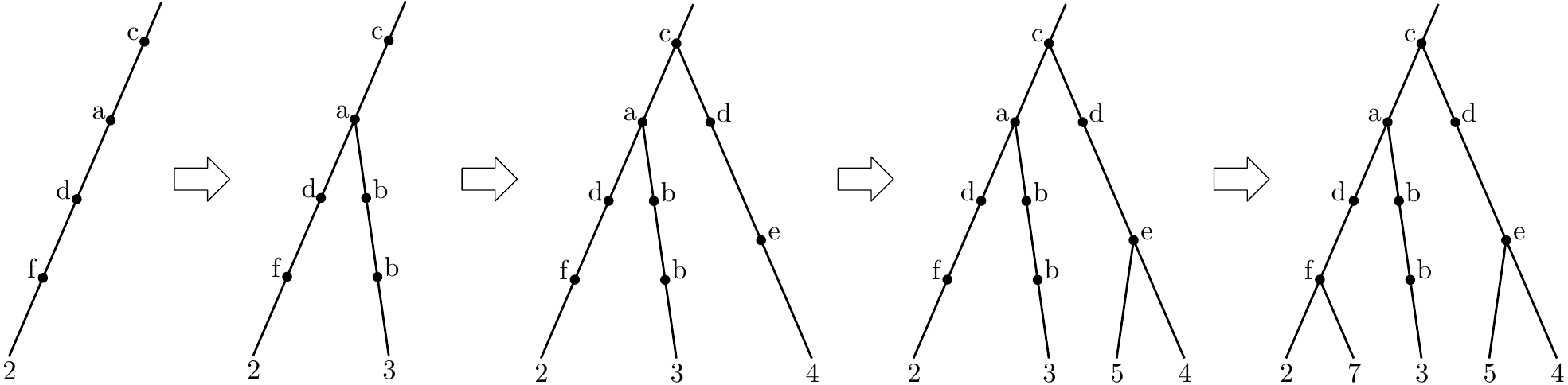}
\end{center}
\caption{The tree constructed from the subwords of the word $cadfabbacdbeef$ and the tuple $(6,1)$. (Thus, $n=7$ and $k=2$.)}\label{tree-const}
\end{figure}

\vspace*{0.35cm} Here, we explain the inverse map of the above map. Therefore, assume that we have given a word from $\mathcal{H}_{n-1,k}$ and an $k$-tuple from $\{1,\ldots,n\}^{\underline{k}}$. We again use the example above, thus, the word is $cadfabbacdbeef$ and the tuple is $(6,1)$.

For the word, we separate it into a word $\omega$ and a set $A$ as in Lemma~\ref{hnk} (for our example: $\omega=cadfabbcdeef$ and $A=\{b,d\}$). Then, we use the second occurrence of the letters from $B:=\{\omega_1,\ldots,\omega_{n-1}\}\setminus A$ (in the example $B=\{a,c,e,f\}$) to separate $\omega$ into $n-k$ subwords: the first subword is from the beginning to the letter before the second occurrence of the first letter from $B$ (in the example this is $cadf$); the next is from the second occurrence of the first letter from $B$ to the letter before the second occurrence of the second letter from $B$ (in the example this is $abb$); etc. For the example this gives the subwords: $cadf, abb, cde, e, f$.

These subwords correspond to nodes along paths in a tree that is successively constructed as follows: the first subword corresponds to the first path which ends in a leaf which is labeled by the smallest integer from $\{1,\ldots,n\}$ which is not in the $k$-tuple (in the example this is $2$); the next subword starts with a letter which is contained on the first path; starting from this letter, we construct a path with labels from the second subword which ends in a leaf which is labeled with the second smallest integer from $\{1,\ldots,n\}$ which is not in the $k$-tuple (in the example this is $3$); etc.; see Figure~\ref{tree-const} for the construction of this tree.

\begin{figure}
\begin{center}
\includegraphics[scale=0.8]{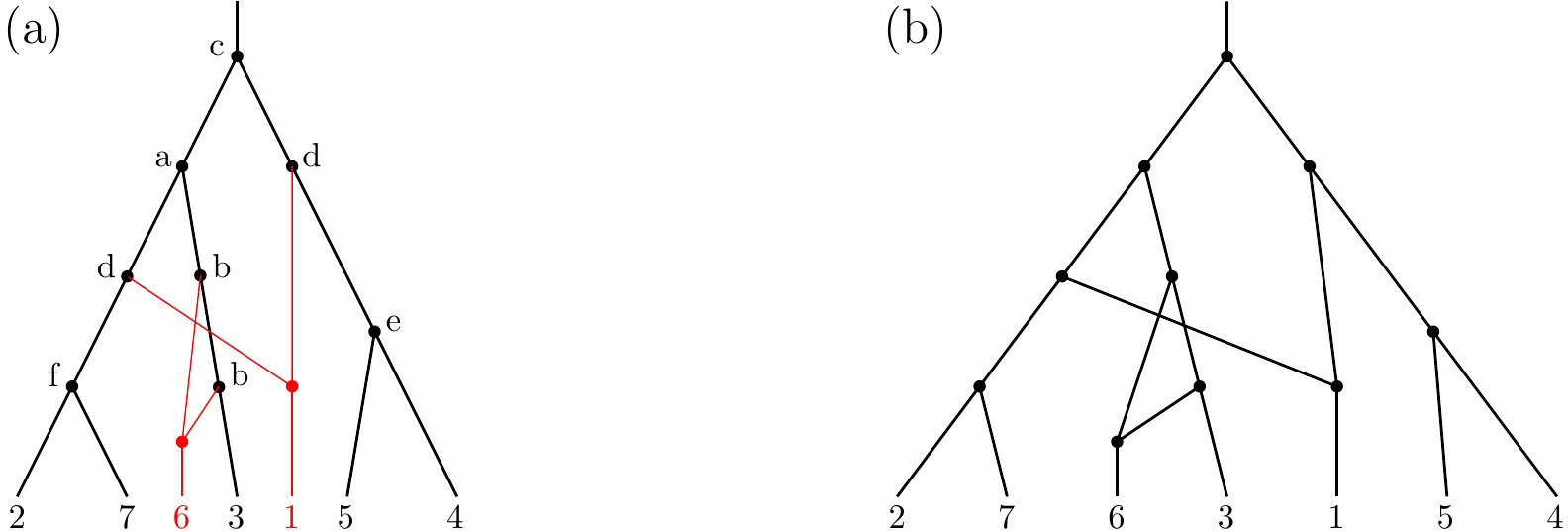}
\end{center}
\caption{(a) The two reticulation nodes are added and they are labeled with the given tuple; (b) Removing all labels of the internal nodes gives the same OTC with which we started in Figure~\ref{path-dec}-(a). (Up to its embedding into the plane.)}\label{final-steps}
\end{figure}

Next, reticulation nodes are attached whose parents are the unary nodes on the tree with identical labels. Moreover, the leaves of the reticulation nodes are labeled with the integers from the $k$-tuple in the order of the second occurrence of the letters of the parents in $\omega$; see Figure~\ref{final-steps}-(a).

Finally, removing all labels of the internal nodes gives the preimage of the map from above; see Figure~\ref{final-steps}-(b).

\section{Final Remarks}

We give a consequence of Theorem~\ref{main-result} and compare our map with the map from \cite{PoBa} ($k=0$) and the restriction of the map of \cite{FuYuZh} ($k=n-1$) to OTCs.

In order to state the consequence, we first give an easy lemma.

\begin{lemma}
We have,
\[
\vert H_{n,k}\vert=\binom{n}{k}(2n-1)!!.
\]
\end{lemma}
\pf This follows directly from Lemma~\ref{hnk} and (\ref{result-bp}) which shows that $\vert C_{n-1,0}\vert$ is the number of phylogenetic trees which is well-known to equal $(2n-3)!!$; see Corollary 2.2.4 in \cite{SeSt}.

Alternatively, $\vert\mathcal{C}_{n,0}\vert$ can also be counted directly. Therefore, recall that the second occurrence of letters of words in $\mathcal{C}_{n,0}$ must occcur in the order $\omega_1,\ldots,\omega_n$. Now, replacing the letters by any permutation gives exactly once all words of length $2n$ with letters from $\{\omega_1,\ldots,\omega_n\}$ where each letter occurs exactly twice. The number of the latter words is $(2n)!/2^n$. This gives
\[
\vert\mathcal{C}_{n,0}\vert=\frac{(2n)!}{2^nn!}=(2n-1)!!
\]
as claimed.\qed

From Theorem~\ref{main-result}, we obtain now the following consequence.

\begin{corollary}
The number of OTCs with $n$ leaves and $k$ reticulation nodes equals
\[
\vert\mathcal{OTC}_{n,k}\vert=\frac{n!}{(n-k)!}\binom{n-1}{k}(2n-3)!!.
\]
\end{corollary}

\begin{remark} This formula was first obtained in \cite{CaZh}.
\end{remark}

With the above formula, it was proved in \cite{FuYuZh} that the number of reticulation nodes of an OTC which is chosen uniformly at random from the set of all OTCs with $n$ leaves satisfies a central limit theorem. (This was not stated as a theorem in \cite{FuYuZh} but is implicit in the proof of Theorem~3 of this paper where even a local limit theorem was proved.)

\begin{theorem}
Denote by $R_n$ the number of reticulation nodes of an OTC which is chosen uniformly at random from all OTCs with $n$ leaves. Then,
\[
\frac{R_n-{\mathbb E}(R_n)}{\sqrt{{\rm Var}(R_n)}}\stackrel{d}{\longrightarrow} N(0,1).
\]
\end{theorem}

We next discuss how our bijective map compares to the ones from \cite{PoBa} and \cite{FuYuZh} for $k=0$ and $k=n-1$, respectively. First, the one from \cite{PoBa} for $k=0$ coincides with our map. (Note that in \cite{PoBa}, no tuple was considered which is legitimate since in the case $k=0$ the tuple is empty.) However, the restriction of the map from $\cite{FuYuZh}$ for $k=n-1$ to OTCs is different. More precisely, note that for the word arising as image of the map of this paper for $k=n-1$, the third occurrence of $\omega_j$ is right after the second occurrence. On the other hand, for the word arising from the map from \cite{FuYuZh} for an OTC, the third occurrences are all at the end, i.e., the prefix of any word arising as image of an OTC under that map is $\omega_1\cdots\omega_{n-1}$.

Finally, the most obvious open question which is raised by this short note is the following: can our bijection be extended to general tree-child networks? Since one-component tree-child networks are still relatively simple and more general tree-child networks can have a much more complicated structure, the answer to this question is, at present, still unclear.

\section*{Acknowledgments}

We thank Louxin Zhang for posing the question to us which was solved in this note. We also thank Yu-Sheng Chang and Tsan-Cheng Yu for joining the discussion about this research. MF was partially supported by the MOST (Ministry of Science and Technology, Taiwan) grant MOST-109-2115-M004-003-MY2; GRY by the MOST grant MOST-110-2115-M-017-003-MY3.

\end{document}